\documentclass[twocolumn,aps,showpacs,amsmath,amssymb,superscriptaddress]{revtex4}
\usepackage{graphicx}

\begin{document}
\newcommand {\nc} {\newcommand}
\nc {\beq} {\begin{eqnarray}}
\nc {\eol} {\nonumber \\}
\nc {\eeq} {\end{eqnarray}}
\nc {\eeqn} [1] {\label{#1} \end{eqnarray}}
\nc {\eoln} [1] {\label{#1} \\}
\nc {\ve} [1] {\mbox{\boldmath $#1$}}

\title{Updated three-body model of $^6$He $\beta$ decay into the $\alpha$ + $d$ continuum}

\author{E.M. Tursunov}
\email{tursune@inp.uz} \affiliation{Institute of Nuclear Physics,
Uzbekistan Academy of Sciences, 100214, Ulugbek, Tashkent,
Uzbekistan}
\author{D. Baye}
\email{dbaye@ulb.ac.be}
\affiliation{Physique Quantique, C.P. 229,
Universit\'e Libre de Bruxelles, B 1050 Brussels, Belgium}
\affiliation{Physique Nucl\'eaire Th\'eorique et Physique Math\'ematique, C.P. 229,
Universit\'e Libre de Bruxelles, B 1050 Brussels, Belgium}
\author{P. Descouvemont}
\email{pdesc@ulb.ac.be}
\affiliation{Physique Nucl\'eaire Th\'eorique et Physique Math\'ematique, C.P. 229,
Universit\'e Libre de Bruxelles, B 1050 Brussels, Belgium}

%\preprint{Version 2}

\date{\today}
\begin{abstract}
The $\beta$-decay process of the $^6$He halo nucleus 
into the $\alpha+d$ continuum is studied in an updated three-body model. 
The $^6$He nucleus is described as an $\alpha+n+n$ system in 
hyperspherical coordinates on a Lagrange-mesh. 
The shape and absolute values of the transition probability per time and energy 
units of new experiments are reproduced with a modified $\alpha+d$ potential. 
The obtained total transition probabilities are $2.48 \times 10^{-6}$ s$^{-1}$ 
for the full energy region and $2.40 \times 10^{-6}$ s$^{-1}$ 
for the cut-off $E>150$ keV. 
The strong cancellation between the internal and halo parts of the 
$\beta$ decay matrix element is a challenge for future {\it ab initio} calculations.  
\end{abstract}
\maketitle
%\begin{multicols}{2}
%\narrowtext
%
The $\beta$-delayed deuteron decay of $^6$He, {\it i.e.} the $\beta$ decay 
of $^6$He into $^4$He and a deuteron, 
\beq
^6{\rm He} \, \to  \, \alpha \, + \, d\,+ e^- \,+ \, \bar{\nu}_e,
\eeqn{eq1}
has been measured several times with various results 
for its very small branching ratio \cite{rii90,bor93,ABB02,raabe09,marek15}. 
The smallness of the branching was first explained as a cancellation 
between the internal and halo parts of the Gamow-Teller matrix element 
by a semi-microscopic model in Ref.~\cite{BSD94}. 
This interpretation was confirmed by later models 
(see references in Ref.~\cite{tur06}) but all results are very sensitive 
to tiny details. 
The branching ratio that we obtained in a three-body model \cite{tur06,tur06E} 
agreed with the data of the most recent experiment at that time \cite{ABB02}. 
This is due to a good description of the ground-state energy and halo of $^6$He 
with an $\alpha+n+n$ wave function and to a potential fitting the $\alpha+d$ 
$s$-wave phase shift and the normalization of the experimental curve. 

Since the publication of our calculation \cite{tur06,tur06E}, two measurements 
\cite{raabe09,marek15} were performed, which update the experimental data 
and challenge our theoretical transition probabilities of the process. 
The first measurement by the ISOLDE collaboration in 2009 \cite{raabe09} 
used the technique of implantation into a highly segmented silicon detector. 
A branching ratio $B = (1.65 \pm 0.10) \times 10^{-6}$ was obtained 
for deuterons with energies above 350 keV with a $6 \%$ error. 
The corresponding transition probability is 
$W = (1.42 \pm 0.09) \times 10^{-6}$ s$^{-1}$ for $E_d > 350$ keV. 
The data slightly underestimates our previous theoretical results, 
$2.04 \times 10^{-6}$ s$^{-1}$ for the full energy range or 
$1.59 \times 10^{-6}$ s$^{-1}$ for a cutoff $E> 370$ keV, 
which agreed with the old experimental results \cite{ABB02}. 

The second measurement \cite{marek15} was performed in 2015 
by the same collaboration at the REX-ISOLDE facility. 
The $^6$He ions were implanted into the optical time projection chamber, 
where the decays with emission of charged particles were recorded. 
This technique allowed the authors to measure the spectrum down to 150 keV 
in the $\alpha + d$ center-of-mass frame. 
The branching ratio for this process amounts to 
$[2.78 \pm 0.07$(stat)$\pm 0.17$(sys)]$\times 10^{-6}$. 
The shape of the spectrum is found to be in a good agreement with the three-body model 
\cite{tur06,tur06E}, while the total transition probability is 
$[2.39 \pm 0.06$(stat)$\pm 0.15$(sys)]$\times 10^{-6}$ s$^{-1}$ 
which is about $20\%$ larger than our theoretical prediction of Ref.~\cite{tur06E}, 
while the shape of the spectrum is in excellent agreement with theory. 
The aim of the present report is to update the theoretical model \cite{tur06,tur06E} 
and describe the new experimental data \cite{marek15} with high precision. 
We also discuss expectations for theoretical progress. 

The $^6$He nucleus is described as an $\alpha+n+n$ system in 
hyperspherical coordinates on a Lagrange mesh (see Ref.~\cite{DDB03} for details). 
The ground-state wave function $\Psi_{^6{\rm He}}(\ve{r},\ve{R})$
is then expressed and normalized in Jacobi coordinates: 
$\ve{r}$ between the neutrons and $\ve{R}$ between the $\alpha$ core 
and the center of mass of these neutrons. 
The transition probability per time and energy units is given by \cite{BD-88}
\beq
\frac{dW}{dE}= \frac{m_e c^2}{\pi^4 v \hbar^2} G_{\beta}^2 f(Q-E) B_{\rm GT}(E),
\eeqn{eq2}
where $m_e$ is the electron mass, $v$ and $E$ are the relative velocity and energy in the center 
of mass frame of $\alpha$ and deuteron, and $G_{\beta}$ is the dimensionless $\beta$-decay constant. 
The Fermi integral $f(Q-E)$ depends on the total kinetic energy $Q-E$ of the electron and antineutrino. 
The mass difference $Q$ is 2.03 MeV. 
The Gamow-Teller reduced transition probability reads 
\beq
B_{\rm GT}(E) = 6\lambda^2 [I_E(\infty)]^2
\eeqn{eq3}
where $\lambda$ is the ratio of the axial-vector to vector coupling constants 
and $I_E(R)$ is the integral \cite{tur06} 
\beq
I_E(R) = \int_0^R u_{\rm eff}(R') u_E(R') dR'.
\eeqn{eq4}
The effective function
\beq 
u_{\rm eff}(R)= R \int_0^{\infty} \psi(r,R) u_d(r) r dr
\eeqn{eq5}
is the overlap of the  $l_x=l_y=L=S=0$ component 
\beq
\psi(r,R) = \langle [[Y_0(\Omega_R) \otimes Y_0(\Omega_r)]_0 \otimes \chi_0]^{00} 
| \Psi_{^6{\rm He}} \rangle,
\eeqn{eq6}
and the $s$-wave radial function of the deuteron $u_d(r)$ 
(see Ref.~\cite{tur06} for details). 
The $l = 0$ scattering wave function $u_E(R)$ is calculated with a simple Gaussian potential 
which reproduces the binding energy of $^6$Li 
and the $\alpha+d$ phase shift $\delta_0$ of the $s$ wave. 
Its asymptotic behavior is $\cos \delta_0 F_0(kr) + \sin \delta_0 G_0(kr)$, 
where $k$ is the wavenumber 
and $F_0$ and $G_0$ are the $l=0$ regular and irregular Coulomb wave functions. 

\begin{figure}[thb]
\setlength{\unitlength}{1 mm}
\begin{picture}(90,55) (0,5) 
\put(0,0){\includegraphics[width=0.5\textwidth]{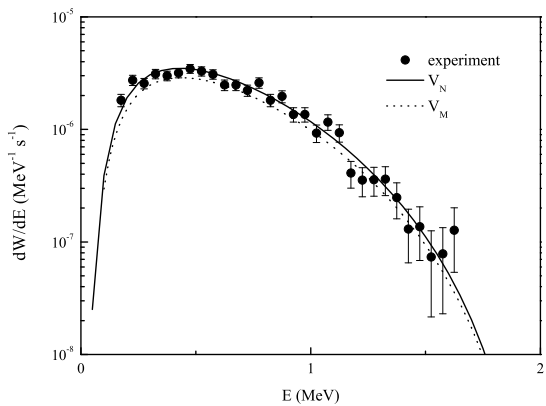}}
\end{picture} \\
\caption{Transition probability per time and energy units $dW/dE$ of
the $^6$He $\beta$ decay into the $\alpha+d$ continuum for the new 
($V_N$, full line) and previous ($V_M$, dotted line) $\alpha+d$ potentials. 
The experimental data are from Ref.~\cite{marek15}.}
\label{Fig1}
\end{figure}
The new data \cite{marek15} can be described by a refitted $\alpha+d$ potential. 
We slightly modify the Gaussian potential $V_M(r)=-79.4\,\exp(-0.21\,r^2)$ 
from Ref.~\cite{tur06E} into $V_N(r)=-80.55\,\exp(-0.2135\,r^2)$ 
which describes equally well the binding energy 1.474 MeV of the $^6$Li ground state 
and the $s$-wave phase shift of the $\alpha+d$ scattering up to 4 MeV,
an energy exceeding the threshold energy 2.03 MeV of the $\beta$ decay. 
Both potentials possess a bound state below the $^6$Li ground state 
which simulates a Pauli forbidden state in the $s$ wave. 
In Fig.~\ref{Fig1}, the transition probabilities per time and energy units $dW/dE$ 
of the $^6$He $\beta$ decay into the $\alpha+d$ continuum for potentials $V_M$ 
and $V_N$ are displayed in comparison with the experimental data from Ref.~\cite{marek15}. 
A three-body hyperspherical wave function with hypermomentum components up to $K=24$ is used. 
As can be seen from the figure, the modified potential 
$V_N$ describes the new data \cite{marek15} pretty well. 
The total transition probability with the $V_N$ potential is 
estimated as $W=2.48 \times 10^{-6}$ s$^{-1}$, 
while for the cut-off $E>150$ keV we obtain 
$W[E>150\ \mathrm{keV}]=2.40 \times 10^{-6}$ s$^{-1}$, 
which is very consistent with Ref.~\cite{marek15}. 

\begin{figure}[thb]
\setlength{\unitlength}{1 mm}
\begin{picture}(90,50) (0,40) 
\put(-5,0){\includegraphics[width=0.5\textwidth]{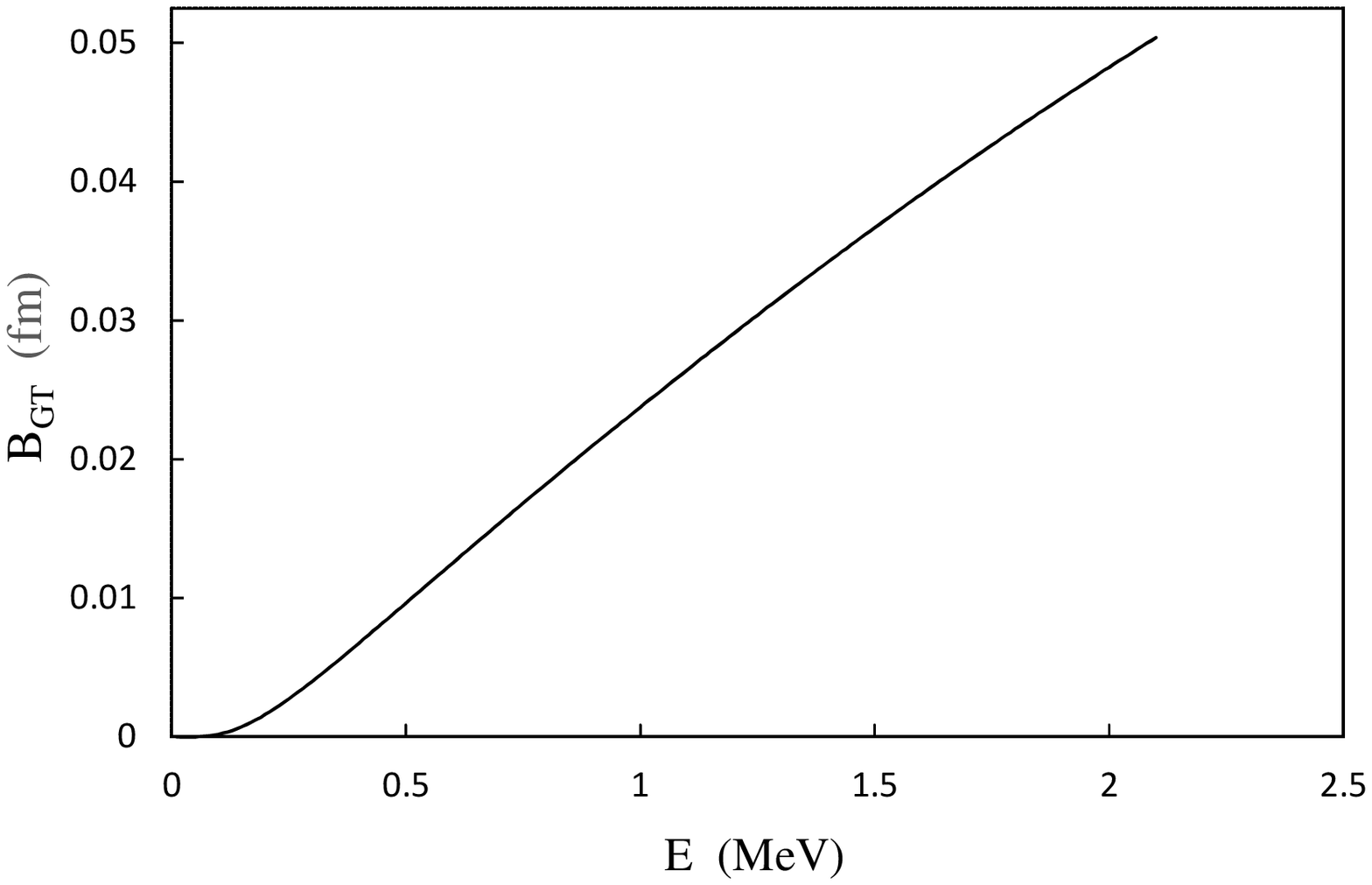}}
\end{picture} \\
\caption{Gamow-Teller reduced transition probability.}
\label{Fig2}
\end{figure}

The shape of the theoretical curve agrees with the new data at low deuteron energies. 
This agreement over an extended energy range again confirms 
the cancellation mechanism of the internal and halo parts 
since it can reproduce both the order of magnitude and energy dependence of the data. 
The Gamow-Teller reduced transition probability is depicted in Fig.~\ref{Fig2} as a function of the energy. 
It is very small under the Coulomb barrier, which explains the fast decrease of the transition 
probability at low energies in Fig.~\ref{Fig1}. 
Above 0.1 MeV, it increases almost linearly. 
The decrease of the transition probability above 0.5 MeV is entirely due to the phase-space factor. 
\begin{figure}[thb]
\setlength{\unitlength}{1 mm}
\begin{picture}(90,50) (0,5) 
\put(0,0){\includegraphics[width=0.5\textwidth]{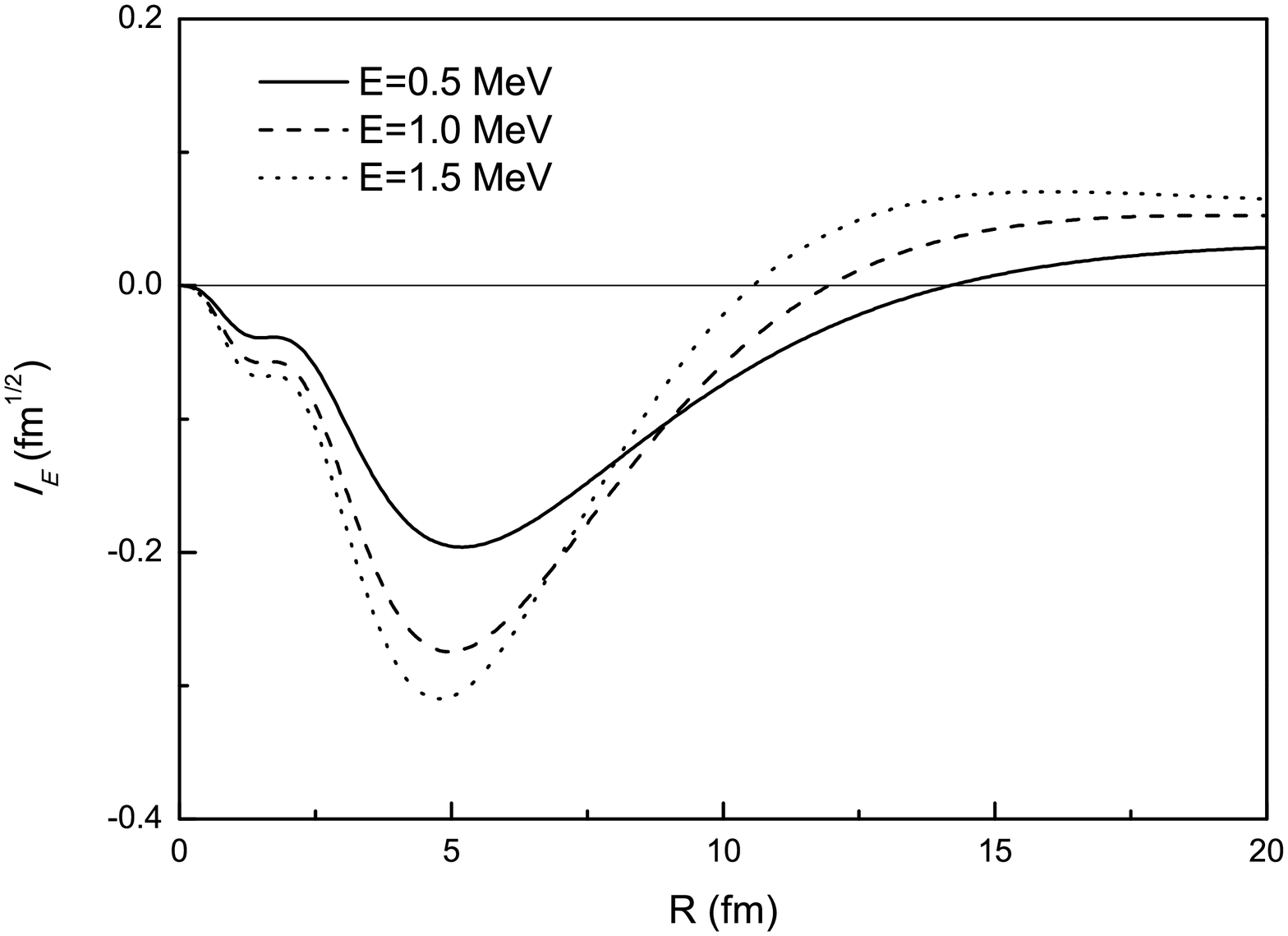}}
\end{picture} \\
\caption{Integrals $I_E(R)$ at $E=0.5$ (full line), 1 (dashed line), 
and 1.5 MeV (dotted line) calculated with the $V_N$ potential.}.
\label{Fig3}
\end{figure}
The shape of the reduced transition probability $B_{\rm GT}$ can be understood with Fig.~\ref{Fig3}. 
Though integral $I_E(R)$ is only observable asymptotically when $R$ tends to infinity, 
its shape contains important physical information about the cancellation mechanism. 
In Fig.~\ref{Fig3}, this integral is represented at three energies: $E = 0.5$, 1, and 1.5 MeV. 
Its absolute value reaches a maximum near 5 fm before a decrease due to 
a change of sign of the $l=0$ scattering wave. 
This decrease continues to large distances because of the large extension of the halo. 
The integral vanishes at a location where the internal and external parts of the integrand exactly cancel each other. 
Beyond this zero of $I_E(R)$, the halo part dominates and the integral changes sign. 
The cancellation mechanism is very sensitive to the location of the $s$-wave node. 
It is stronger at small energies where this node is at a larger distance. 
Hence, $B_{\rm GT}(E)$ progressively increases when this node moves to the left with increasing energy. 
The sensitivity of $B_{\rm GT}(E)$ to the exact location of this node 
will make model-independent {\it quantitative} predictions of the transition probability very difficult. 
In the present model, experimental data on the transition probability are needed 
to fix the phenomenological $\alpha+d$ potential 
which is not constrained enough by the phase shifts. 
If a new experiment leads to a more accurate normalization of these data, 
this potential may have to be refitted. 

The description of the delayed $\beta$ decay is accessible to {\it ab initio} calculations 
since both $^6$He \cite{RQN16} and the $\alpha+d$ scattering \cite{HQN15} have been studied in this way. 
Since these models have no free parameter, their results will be very sensitive 
to the delicate cancellation mechanism and small inaccuracies may lead to large disagreements with experiment. 
Moreover, it is not clear whether these models are able yet to accurately describe 
the halo of $^6$He up to about 20 fm as required by the cancellation mechanism. 
In these models, the energy-dependent integral would be defined by 
\beq
&& I_E(R) = \frac{k}{\sqrt{2\pi}} 
\eol
&& \times \int_0^R 
\langle \Psi^{10+}_{\alpha+d} | \delta(\rho-R') 
\sum_{j=1}^6 t_{j-} s_{jz} | \Psi^{00+}_{^6{\rm He}} \rangle dR',
\eeqn{eq7}
where $\ve{s}_j$ and $\ve{t}_j$ are the dimensionless spin and isospin operators of nucleon $j$ 
and $\ve{\rho}$ is the relative coordinate between the $^4$He and deuteron centers of mass. 
The  six-nucleon wave functions $\Psi^{00+}_{^6{\rm He}}$ and $\Psi^{10+}_{\alpha+d}$ represent 
the $^6$He ground-state and the $J^\pi=1^+$ partial wave of an $\alpha+d$ scattering wave 
normalized asymptotically to $\exp(i\ve{k}\cdot\ve{\rho})$, respectively. 
It is important to verify whether such a microscopic calculation confirms at least 
the internal part of $I_E(R)$ depicted in Fig.~\ref{Fig3}. 
Information about the location of the $s$-wave node would also be essential. 

In conclusion, the three-body model based on the hyperspherical 
Lagrange-mesh method \cite{tur06,tur06E} is updated for the 
description of new experimental data \cite{raabe09,marek15}. 
It is shown that the new data can be described pretty well with the help 
of a modification of the $\alpha+d$ potential in the $s$ wave, 
while keeping the descriptions of the binding energy of the $^6$Li ground state and 
$s$-wave $\alpha+d$ phase shift and, importantly, the presence of a Pauli forbidden state. 
The modification of the potential results in a shift of the nodal position 
of the $l=0$ relative scattering wave function, 
which affects the values of the effective integral for the $\beta$-decay matrix elements. 
Nevertheless, the main conclusion of the three-body model of Ref.~\cite{tur06} remains valid, 
{\it i.e.}\ that the lowering of the $\beta$-decay transition probability occurs 
due to a cancellation effect of the internal and external parts of 
the Gamow-Teller matrix element. 
An important open question is how to fix the potential of the $\alpha+d$ relative motion 
without fitting its parameters to $\beta$-decay data. 
The answer could come from microscopic approaches. 
We suggest that the effective integral $I_E(R)$ should provide an important link 
between partly phenomenological three-body models and {\it ab initio} descriptions. 
\section*{Acknowledgments}
We thank Marek Pf\"utzner for sending us the experimental data. 
E.M.T.\ and P.D.\ acknowledge the support of the Fonds de la Recherche 
Scientifique - FNRS, Belgium. 
\end{document}